\documentclass[journal=jacsat,manuscript=article,layout=traditional]{achemso}

\usepackage{amsmath}
\usepackage{listings}
\usepackage{enumerate}
\usepackage{graphicx}
\usepackage{dcolumn}
\usepackage{bm}
\usepackage{wrapfig}
\graphicspath{{Figures/}}
\usepackage{color}
\usepackage{xr}

\newcommand{\prob}{\mathcal{P}}
\newcommand{\Gammapdf}{\mathcal{G}}
\newcommand{\normal}{\mathcal{N}}
\newcommand{\mtx}[1]{\bm{#1}}

\author{J.~Shepard Bryan IV}
\affiliation{Center for Biological Physics, Department of Physics, Arizona State University}

\author{Ioannis Sgouralis}
\affiliation{Center for Biological Physics, Department of Physics, Arizona State University}

\author{Steve Press{\'e}}
\email{spresse@asu.edu}
\affiliation{Center for Biological Physics, Department of Physics, Arizona State University}
\alsoaffiliation{School of Molecular Sciences, Arizona State University}

\title[Effective forces with Gaussian processes]{Inferring effective forces for Langevin dynamics using Gaussian processes}

\begin{document}

\date{\today}

\begin{abstract}
Effective forces--derived from experimental or {\it in silico} molecular dynamics time traces--are critical in developing reduced and computationally efficient descriptions of otherwise complex dynamical problems. Thus, designing methods to learn effective forces efficiently from time series data is important. Of equal importance is the fact that methods should be suitable in inferring forces for undersampled regions of the phase space where data are limited. Ideally, a method should {\it a priori} be minimally committal as to the shape of the effective force profile, exploit every data point without reducing data quality through any form of binning or pre-processing, and provide full credible intervals (error bars) about the prediction. So far no method satisfies all three criteria. Here we propose a generalization of the Gaussian process (GP), a key tool in Bayesian nonparametric inference and machine learning, to achieve this for the first time.

\end{abstract}

\maketitle

\section{Introduction}

Compressing complex and noisy observations into simpler effective models is the goal of much of biological 
and condensed matter physics~\cite{doi:10.1021/acs.jctc.8b00869,doi:10.1021/acs.jpclett.7b02185,Sgouralis2018,Jazani2019}. Though there are many models of reduced dynamics, in this proof-of-principle work, we focus on the Langevin equation~\cite{reif2009fundamentals,Schlick} to describe the time evolution of a spatial degree of freedom such as a particle's position in physical space or the distance between two marked locations such as protein sites. This equation is governed by deterministic as well as random (thermal) forces dampened by frictional forces~\cite{zwanzig2001nonequilibrium}. 

\begin{figure*}[tbp]
\centering	
\includegraphics[width=0.9\textwidth]{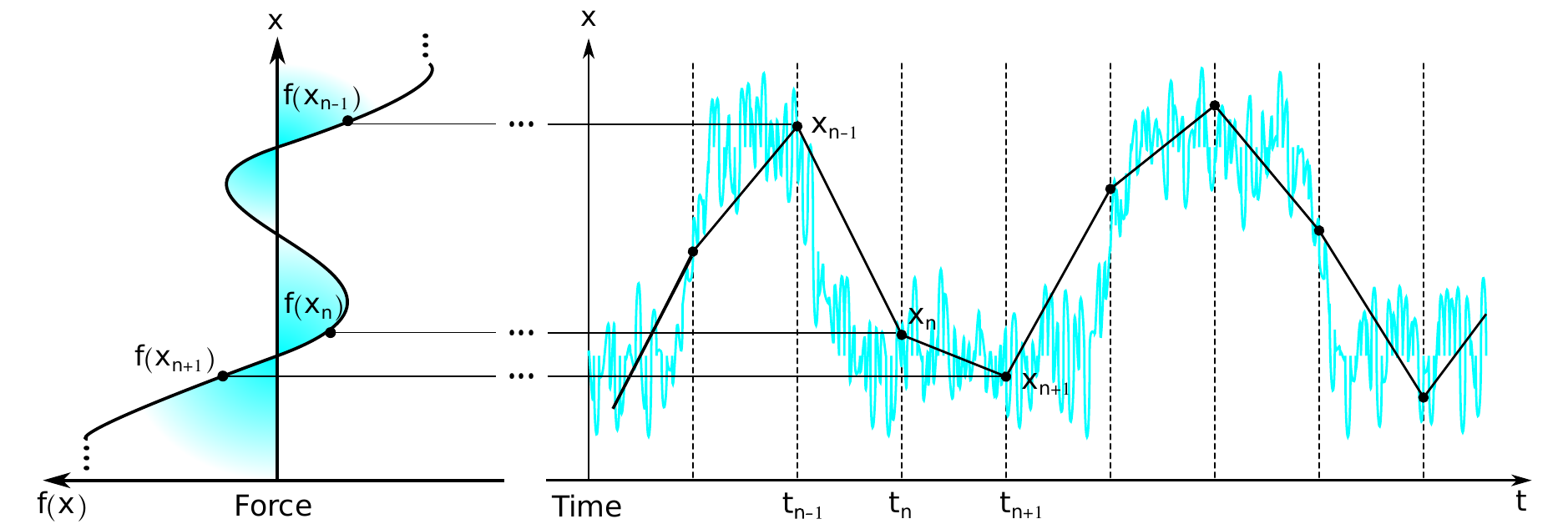}
\caption{\textbf{Illustration of the goal of the analysis presented}. A scalar degree of freedom, $x$, such as a particle's position or an intra-molecular distance, is measured at discrete time levels, $t_n$.  The measurements obtained, $x_n$, in turn, are used to find an effective force $f(x)$. Despite the discrete measurements, the effective force found is a continuous function over $x$ and extends also over positions that may not necessarily coincide with the measured ones $x_n$.}
\label{fig:KillerWOError}
\end{figure*}

Concretely, the ability to infer effective forces (Fig.~\ref{fig:KillerWOError}) from sequences of data, hereonin ``time traces", is especially useful in: 
1) quantitative single molecule analysis~\cite{Manzo2015,Meijering2012,Sgouralis2017,sgouralis2017icon,Presse2014,Presse2013} having revealed many interesting phenomena well modeled within a Langevin framework, for example, confinement~\cite{Kusumi1993,Simson1995,Welsher2015}; and 2) molecular dynamics (MD) simulations where effort has already been invested in approximating exact force fields with coarse-grained potentials~\cite{Wang2018,Poltavsky2017,Kumar1992,Foley2015, Izvekov2020, MARTINI_MD}. The key behind these coarse-grained force fields is to capture the complicated effects of the extra- or intra-molecular environment on the dynamics of a degree of freedom of interest free of atomistic details.

Naively, to obtain an effective force from such a time trace, we would create a histogram for the positions in the time trace ignoring time, take the logarithm of the histogram to find the potential invoking Boltzmann statistics, and then calculate the gradient of the potential~\cite{reif2009fundamentals}. In practice, this naive method requires arbitrary histogram bin size selection, large amounts of data to converge, and incorrectly estimates the effective force in oversampled and undersampled regions~\cite{Turkcan2012}.

Ideally a method to learn effective forces from time series data would be 1) minimally committal, {\it a priori}, to the effective force's profile, 2) provide full credible intervals (error bars) about the prediction, and 3) exploit every data point without down-sampling such as binning or pre-processing. Existing Bayesian methods can be tailored to satisfy the first two criteria above~\cite{Masson2009, Turkcan2012}, however, these methods do not cover the last criteria. 

Existing methods are not free of pre-processing such as spatial segmentation and losing information through averaging the force in each spatial segment~\cite{Turkcan2012,Masson2009}. Neither of these, or similar pre-processing steps, are major limitations when vast amounts of data are available. Yet data is quite often limited in both MD trajectories and experiments on account of limited computational resources ({\it in silico}) or otherwise damaging ({\it in vitro}) or invasive ({\it in vivo}) experimental protocols~\cite{Lee2017}. These limitations highlight once more the importance of exploiting the information encoded in each data point as efficiently as possible.

To develop a method satisfying all three criteria, we use tools within Bayesian nonparametrics, in particular Gaussian processes (GPs)\cite{Rasmussen:2005:GPM:1162254}. Such nonparametrics are required because force is a continuous function and so learning it cannot be achieved through conventional Bayesian statistics~\cite{gelman2013bayesian,volToussant}. We use GPs as priors to learn the effective forces from a time trace. We show that our method outperforms existing methods requiring discretization because our method allows inference over the entire space including unsampled regions. In addition, as we will demonstrate, our method converges to a more accurate force with fewer data points as compared to existing methods.

\section{Methods}

Here we set up both the dynamical (Langevin) model and discuss the inference strategy to learn the effective force from data. As with all methods within the Bayesian paradigm~\cite{volToussant,gelman2013bayesian,Lee2017,Sgouralis2018,Tavakoli2017}, we provide not only a point estimate for the maximum {\it a posteriori} (MAP) effective force estimate, but also achieve full posterior inference with credible intervals.

\subsection{Dynamics model}
\label{Dynamics model}

For concreteness, here we start with one dimensional over-damped Langevin dynamics given by\cite{zwanzig2001nonequilibrium}
\begin{align}
\zeta \dot{x} &= f(x) + r(t)\label{langevin}
\end{align}
where $x(t)$ is the position coordinate at time $t$; $f(x)$ is the effective force at position $x$; and $\zeta$ is the friction coefficient. The thermal force, $r(t)$, is stochastic and has moments $\langle r(t)\rangle=0$ and $\langle r(t)r(t')\rangle=2\zeta kT\delta(t-t')$, where $\langle\cdot\rangle$ denotes ensemble averages over realizations; $T$ is the temperature and $k$ is Boltzmann's constant.

The positional data from the time trace is $\mtx{x}_{1:N}$. These positions, $x_n=x(t_n)$, are indexed with $n$, the time level label, where $n=1,\dots,N$. In this study, we assume the time levels $t_n$ are known. Further, we assume the step size, $\tau=t_{n+1}-t_n$, is constant throughout the entire trace $\mtx x_{1:N}$, although we can easily relax this requirement.

Our assumed model in Eq.~\eqref{langevin} is over-damped and noiseless, meaning that there is no observation error, as would be relevant to MD. The noiseless assumption is valid when observation error, which we do not incorporate in the present formulation, is much smaller than the kicks induced by the thermal force $r(t)$.

\subsection{Inference model}

Given the dynamical model and the time trace, $\mtx{x}_{1:N}$, our goal is to learn the force at each point in space, $f(\cdot)$ (normally, we would write $f(x)$ but this might be misleading as it is unclear if it means the function $f$ or the value of the function evaluated at $x$),  and the friction coefficient, $\zeta$. To achieve this, we use a Bayesian approach (See SI). We propose a zero-mean Gaussian process (GP) prior on $f(\cdot)$ and an independent Gamma prior on $\zeta$, that is $f(\cdot)\sim \mathcal{GP}\left(0,K(\cdot,\cdot)\right)$ and $\zeta\sim\Gammapdf(\alpha,\beta)$.

As the data likelihood (discussed below) assumes a Gaussian form in $f(\cdot)$ for the thermal kicks, the posterior $f(\cdot)|\zeta,\mtx x_{1:N}$ is also a GP by conjugacy. That is, from our GP posterior, we may sample continuous curves $f(\cdot)|\zeta,\mtx x_{1:N}$. To carry out our computations, we evaluate this curve on test points, $\mtx{x}^*_{1:M}$, that may be chosen arbitrarily. Specifically, we evaluate either $\mtx{f^*}_{1:M}$ or $\mtx{f^*}_{1:M}|\zeta,\mtx x_{1:N}$, where $f^*_m=f(x_m^*)$ (See SI). Because the GP allows us to choose the test points $\mtx x^*_{1:M}$ arbitrarily, including their population $M$ and the specific position $x_m^*$ or each one, our $\mtx f^*_{1:M}$ and $\mtx f^*_{1:M}|\zeta,\mtx x_{1:M}^*$ are {\it equivalent} to the entire functions $f(\cdot)$ and $f(\cdot)|\zeta,\mtx x_{1:N}$, respectively. In other words, using the test points $\mtx x_{1:M}^*$ we do not introduce any discretization or any approximation error whatsoever.

\subsubsection{Constructing the likelihood}

The shape of the likelihood is itself dictated by the physics of the thermal kicks, {\it i.e.}, in this case uncorrelated Gaussian kicks. To find the likelihood for a time trace of positions, $\mtx{x}_{1:N}$, we consider the following discretized overdamped Langevin equation under the forward Euler scheme~\cite{Gromacs,Schlick,LeVeque2007}
\begin{align}
\label{eq:FEuler}
\frac{\zeta}{\tau}(x_{n+1}-x_n) &= f\left( x_n \right) + r_{n}
\end{align}
where $r_n\sim\normal\left(0,2\zeta kT/\tau\right)$. Since $r_{n}$ is a Normal random variable, the change in position $x_{n+1}-x_n|x_n$ is also a Normal random variable. In fact, rearranging Eq.~\eqref{eq:FEuler} we obtain
\begin{align}
    x_{n+1}|f(\cdot),\zeta,x_{n} &\sim \normal\left( \frac{\tau}{\zeta} f\left(x_n\right) + x_{n}, \frac{2\tau kT}{\zeta} \right).
\end{align}
Assuming a known initial position, $x_1$, the likelihood of the rest of the time trace is then the product of each of these Normals
\begin{align}
    \prob\left(\mtx{x}_{2:N}|f(\cdot), \zeta, x_1 \right) &= \prod_{n=2}^N \normal\left(x_{n};  \frac{\tau}{\zeta}f\left(x_{n-1}\right) + x_{n-1},\frac{2\tau kT}{\zeta} \right) \label{likelihoodlang}
\end{align}
which may be combined into a single multivariate one
\begin{align}
    \prob\left(\mtx{x}_{2:N}|f(\cdot), \zeta, x_1 \right) &= \normal_{N-1}\left( \tau \mtx{v}_{1:N-1}; \frac{\tau}{\zeta}\mtx{f}_{1:N-1}, \frac{2\tau kT}{\zeta}\mtx{I}  \right) \label{likelihood1}
\end{align}
where $v_n = (x_{n+1}-x_n)/\tau$, $f_n = f\left(x_n\right)$, $\mtx{I}$ is the $(N-1)\times(N-1)$ identity matrix, and $\mtx{f}_{i:j}$ regroups all values of $f_n$ from $n=i$ to $n=j$, similarly for $\mtx{v}_{i:j}$.

\subsubsection{GP prior and posterior for force}

For our force we choose a zero mean Gaussian process prior $f(\cdot)\sim \mathcal{GP}\left(0,K(\cdot,\cdot)\right)$ with a kernel, $K(\cdot,\cdot)$, assuming the familiar squared exponential form~\cite{Rasmussen:2005:GPM:1162254}
\begin{align}
    K(x,x')&=\sigma^2\exp\left(-\frac{1}{2}\left(\frac{x-x'}{\ell}\right)^2\right). \label{hyperstuff}
\end{align}
The discretized GP then assumes the form
\begin{align}
\prob\left(\mtx{f}_{1:N-1}, \mtx{f}^*{_{1:M}}\right) &=
        \normal_{N+M-1}\left(\left[\begin{matrix} \mtx{f}_{1:N-1}\\ \mtx{f}^*{_{1:M}}\end{matrix} \right] ; \mtx{0}, \left[ \begin{matrix} \mtx{K} & \mtx{K_*} \\ \mtx{K_*}^T & \mtx{K_{**}} \end{matrix} \right]\right) \label{gpprior}
\end{align}
where the covaraince matrices $\mtx K,\mtx K_*,\mtx K_{**}$ are given in the SI. For this choice of prior (Eq.~\eqref{gpprior}) and likelihood (Eq.~\eqref{likelihood1}), the posterior distribution of the effective force given friction coefficient and data is
\begin{align*}
        \prob\left(\mtx{f}_{1:N-1}, \mtx{f}^*{_{1:M}}| \mtx{x}_{1:N},\zeta \right) &\propto \prob\left( \mtx{x}_{2:N} | \mtx{f}_{1:N-1}, \mtx{f}^*{_{1:M}},\zeta,x_1 \right) \prob\left( \mtx{f}_{1:N-1},\mtx{f}^*{_{1:M}}| \zeta,x_1 \right)
\end{align*}
which simplifies to
\begin{align}
        \prob\left(\mtx{f}_{1:N-1}, \mtx{f}^*{_{1:M}}, |\mtx{x}_{1:N},\zeta\right) &\propto \prob\left( \mtx{x}_{2:N} | \mtx{f}_{1:N-1},\zeta,x_1 \right) \prob\left( \mtx{f}_{1:N-1}, \mtx{f}^*{_{1:M}}\right).\label{likelihood2}
\end{align}
Rewriting Eq.~\eqref{likelihood1}, as
\begin{align}
    \prob\left(\mtx{x}_{2:N}|f(\cdot), \zeta, x_1 \right) &\propto \normal_{N+M-1}\left( \left[\begin{matrix} \mtx{f}_{1:N-1}\\ \mtx{f}^*{_{1:M}} \end{matrix}\right]; \left[\begin{matrix} \zeta \mtx{v}_{1:N-1}\\ \mtx{0} \end{matrix}\right], \left[\begin{matrix} \frac{2\zeta kT}{\tau}\mtx{I} &\mtx{0}\\\mtx{0}&\epsilon\mtx{I} \end{matrix}\right]  \right)
\end{align}
we combine~\cite{matrixcookbook} the Gaussians in Eq.~\eqref{likelihood2} and then take the limit as $\epsilon$ goes to zero. After this we get
\begin{align}
\prob\left(\mtx{f}_{1:N-1},\mtx{f}^*{_{1:M}}|\mtx{x}_{1:N},\zeta \right) &\propto
        \normal_{N+M-1}\left(\left[\begin{matrix} \mtx{f}_{1:N-1}\\ \mtx{f}^*{_{1:M}}\end{matrix} \right] ;\left[\begin{matrix} \zeta \mtx{v}_{1:N-1}\\ \mtx{0} \end{matrix}\right],\left[ \begin{matrix} \mtx{K}+\frac{2\zeta kT}{\tau}\mtx{I} & \mtx{K_*} \\ \mtx{K_*}^T & \mtx{K_{**}} \end{matrix} \right]\right).
\end{align}
The marginal distribution for $\mtx{f}^*{_{1:M}}$ in this setup is~\cite{Bishop}
\begin{align}
    \prob(\mtx{f}^*_{1:M}|\mtx{x}_{1:N},\zeta) &\propto \normal_{M}\left(\mtx{f}^*_{1:M};   \tilde{\mtx{\mu}},  \tilde{\mtx{K}}\right).
    \label{Fposterior}
\end{align}
where
\begin{align}
    \tilde{\mtx{\mu}} &= \zeta \mtx{K}_*^T\left(\mtx{K} + \frac{2\zeta kT}{\tau} \mtx{I}\right)^{-1}  \mtx{v}_{1:N-1}\\
    \tilde{\mtx{K}} &= \mtx{K}_{**} - \mtx{K}_*^T \left(\mtx{K} + \frac{2\zeta kT}{\tau} \mtx{I}\right)^{-1} \mtx{K}_*.
\end{align}
Further details of this derivation can be found in~\citet{Rasmussen:2005:GPM:1162254} and~\citet{Do2007}.
The MAP estimate for $f(\cdot)|\zeta,\mtx{x}_{1:N}$ evaluated at the test points is provided by $\tilde{\mtx{\mu}}$ while one standard deviation around each such effective force estimate (error bars) is provided by the square root of the corresponding diagonal entry of the covariance matrix, for example, one standard deviation around $\tilde{\mu}_{17}$ is $\sqrt{\tilde{K}_{17,17}}$. To find the effective potential from here, we numerically integrated the effective force over the test points (see SI).

\subsection{Gibbs sampling for friction coefficient}

We now relax the assumption that $\zeta$ is known. Choosing a Gamma prior over $\zeta$ gives the marginal posterior
\begin{align}
    \prob\left(\zeta|\mtx{x}_{1:N},\mtx{f}_{1:N-1}\right) &\propto \normal_{N-1}\left( \tau \mtx{v}_{1:N-1}; \frac{\tau}{\zeta}\mtx{f}_{1:N-1}, \frac{2\tau kT}{\zeta}\mtx{I}  \right) \times \Gammapdf\left( \zeta | \alpha, \beta\right) \label{zetapdf}
\end{align}
where, after the proportionality, we keep track of all terms dependent of $\zeta$. Note that we can sample $\zeta$ from this distribution using the proportionality in Eq.~\eqref{zetapdf} without any need to keep track of the terms that do not depend on $\zeta$~\cite{Bishop}. That is, we can proceed with our inference without any need to normalize Eq.~\eqref{zetapdf}. 

As we now have explicit forms for $\prob\left(\zeta|\mtx{x}_{1:N},\mtx{f}_{1:N-1}\right)$ and  $\prob\left( \mtx{f}_{1:N-1} | \zeta,\mtx{x}_{1:N} \right)$, we learn both $\mtx{f}$ and $\zeta$ using a Gibbs sampling scheme~\cite{GibbsSampling} which starts with an initial value for both $\mtx{f}_{1:N-1}$ and $\zeta$. We outline this algorithm here:
\begin{itemize}
    \item Step 1: Choose initial $\zeta^{(0)}, \mtx{f}_{1:N-1}^{(0)}$
    
    \item Step 2: For many iterations, $i$:
    
    \begin{itemize}
    \item Sample a new force given the previous friction coefficient, $\mtx{f}_{1:N-1}^{(i)} \sim \prob(\mtx{f}_{1:N-1}| \zeta^{(i-1)},\mtx{x}_{1:N})$.

    \item Sample a new friction coefficient given the new force, $\zeta^{(i)} \sim \prob(\zeta| \mtx{x}_{1:N},\mtx{f}_{1:N-1}^{(i)})$.
    \end{itemize}
\end{itemize}
To sample the force we use Eq.~\eqref{Fposterior} with $\mtx x^*_{1:M} = \mtx x_{1:N-1}$. To sample the friction coefficient we must use a Metropolis-Hastings update~\cite{GibbsSampling,volToussant,RobertCasellas}.

This algorithm samples force and friction coefficient pairs $(\mtx{f}_{1:N-1}^{(i)},\zeta^{(i)})$ from the posterior $\prob(\mtx{f}_{1:N-1},\zeta|\mtx{x}_{1:N})$. After many iterations the sampled pairs converge to the true value~\cite{volToussant}. Once we have a large number of pairs, we distinguish the pair which maximizes the posterior. This pair constitutes the MAP friction coefficient and force estimates $( \mtx{\hat f}_{1:N-1},\hat \zeta$).

\section{Results}

To demonstrate our method, we show that we can accurately learn the force from a simple potential while assuming that the friction coefficient, $\zeta$, is known. We show that the accuracy of our method scales with the number of data points. We then demonstrate that our method can learn a more complicated force, while still assuming known friction coefficient. We show that the effectiveness of our method depends on the stiffness, $\zeta/\tau$, of the system. Finally, we allow the friction coefficient to be unknown and use Gibbs sampling to learn both $f(\cdot)$ and $\zeta$ simultaneously.

\subsection{Demonstration}

\begin{figure*}[tbp]
\centering	
\includegraphics[width=0.9\textwidth]{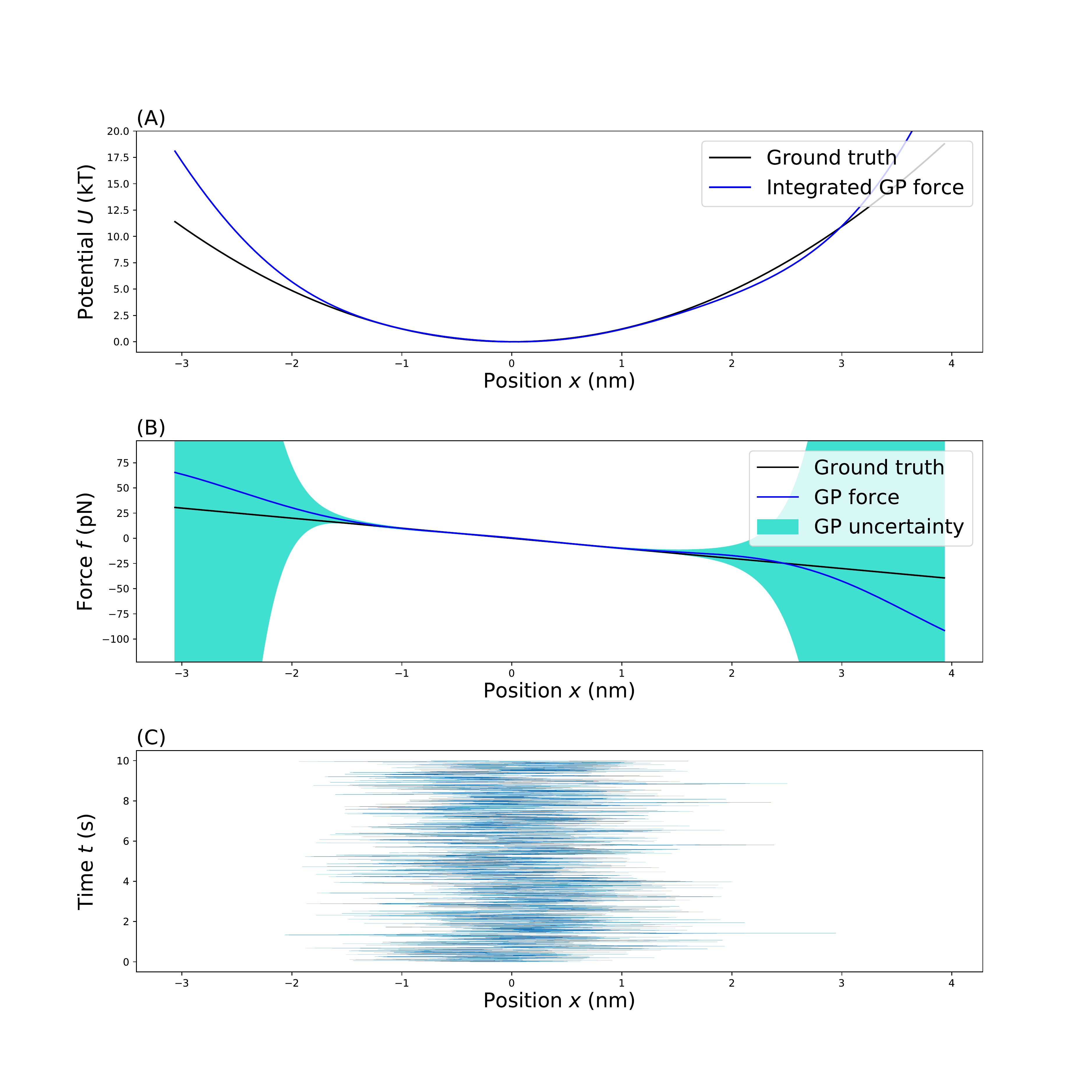}
\caption{\textbf{Testing our method on a harmonic potential}. This plot shows the inference of an effective force (and resulting effective potential) from a simple one-dimensional harmonic well test case. (A) the ground truth potential and our estimate obtained by integrating the MAP effective force estimate, (B) the ground truth force and the GP inferred effective force with uncertainty. (C) the time trace on which inference was performed.}
\label{fig:simplepic1}
\end{figure*}

To demonstrate our method, we applied it to a synthecically generated time trace, consisting of $N=10^4$ data points, of a particle trapped in a harmonic potential well corresponding to the force $f(x)=-Ax$. The values used in the simulation were $\zeta = 100~\mbox{pg/}\mu\mbox{s}$, $T = 300~\mbox{K}$, and $A=10~\mbox{pN/nm}$ probed at $\tau = 1~\mu\mbox{s}$. We looked for the effective force at 500 evenly spaced test points, $\mtx{x}^*_{1:M}$, around the spatial range of the time trace. The hyperparameters used that appear in Eq.~\eqref{hyperstuff} were $\sigma = \alpha\tau(v_{max}-v_{min})$ and $\ell = (x_{max}-x_{min})/2$ with $\alpha = 1~\mbox{pN}\mbox{/nm}$. Briefly, here our motivation for using these values is to set a length, $\ell$, that correlates points over the spatial range covered in the time trace and yet still allows distant regions to be independent, and to set a prefactor, $\sigma$, which gives uncertainty proportional to the range of measured forces. The sensitivity of our method to the particular choice of hyperparameters is analyzed in the SI. 

For simple forces lacking finer detail, the choice of hyperparameters is less critical especially as the number of data points, $N$, increases (See  SI). Similarly, agreement of the predicted effective force with the ground truth is best when the effective force is close to zero ({\it i.e.}, in regions of the effective potential where data points are more commonly sampled). 
As we look further out from the center where there are fewer data points, the uncertainty increases, however, the MAP estimate is still close to the ground truth. This is because the GP prior imposes a smoothness assumption, set by the hyperparameters in Eq.~\eqref{hyperstuff}, which favors maintaining the trend from the MAP estimate from better sampled regions. That is, regions with more data. In regions with no data, the uncertainty eventually grows to the prior value (determined by the first hyperparameter, $\sigma$) and the MAP estimate for $f(\cdot)$ eventually converges to the prior mean ($f(\cdot)=0$).

\begin{figure*}[tbp]
\centering	
\includegraphics[width=0.9\textwidth]{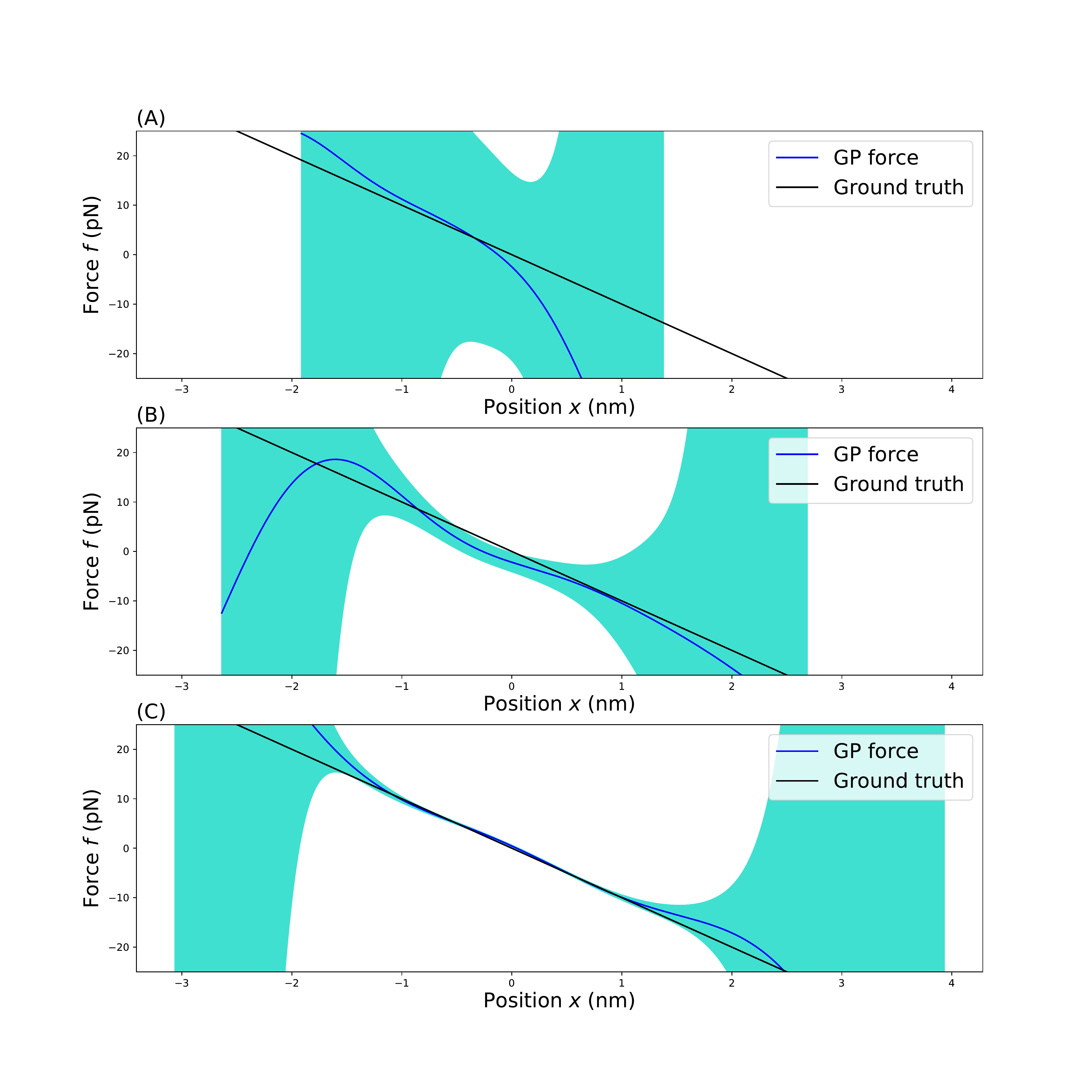}
\caption{\textbf{Testing our method as a function of the number of data points available}. Here we show the effective force inferred from trajectories simulated for a particle trapped in a one dimensional harmonic well with a different number of data points. (A) 100 data points. (B) 1000 data points. (C) 10000 data points. Parameters used are specified in the main body.}
\label{fig:Ndata1}
\end{figure*}

\subsubsection{Varying the number of datapoints}

Fig.~\ref{fig:Ndata1} shows the dependency of the prediction as a function of an increasing number of data points, $N$. Even for a dataset as small as 100 points, the GP gives a reasonable shape for the inferred effective force. However, the uncertainty remains high. As we increase the number of points to 1000 (Fig.~\ref{fig:Ndata1}B) and 10,000 (Fig.~\ref{fig:Ndata1}C) points, we see that the predicted effective force converges to the ground truth force in well-sampled regions with high uncertainty relegated to the undersampled regions.

\begin{figure*}[tbp]
\centering	
\includegraphics[width=0.9\textwidth]{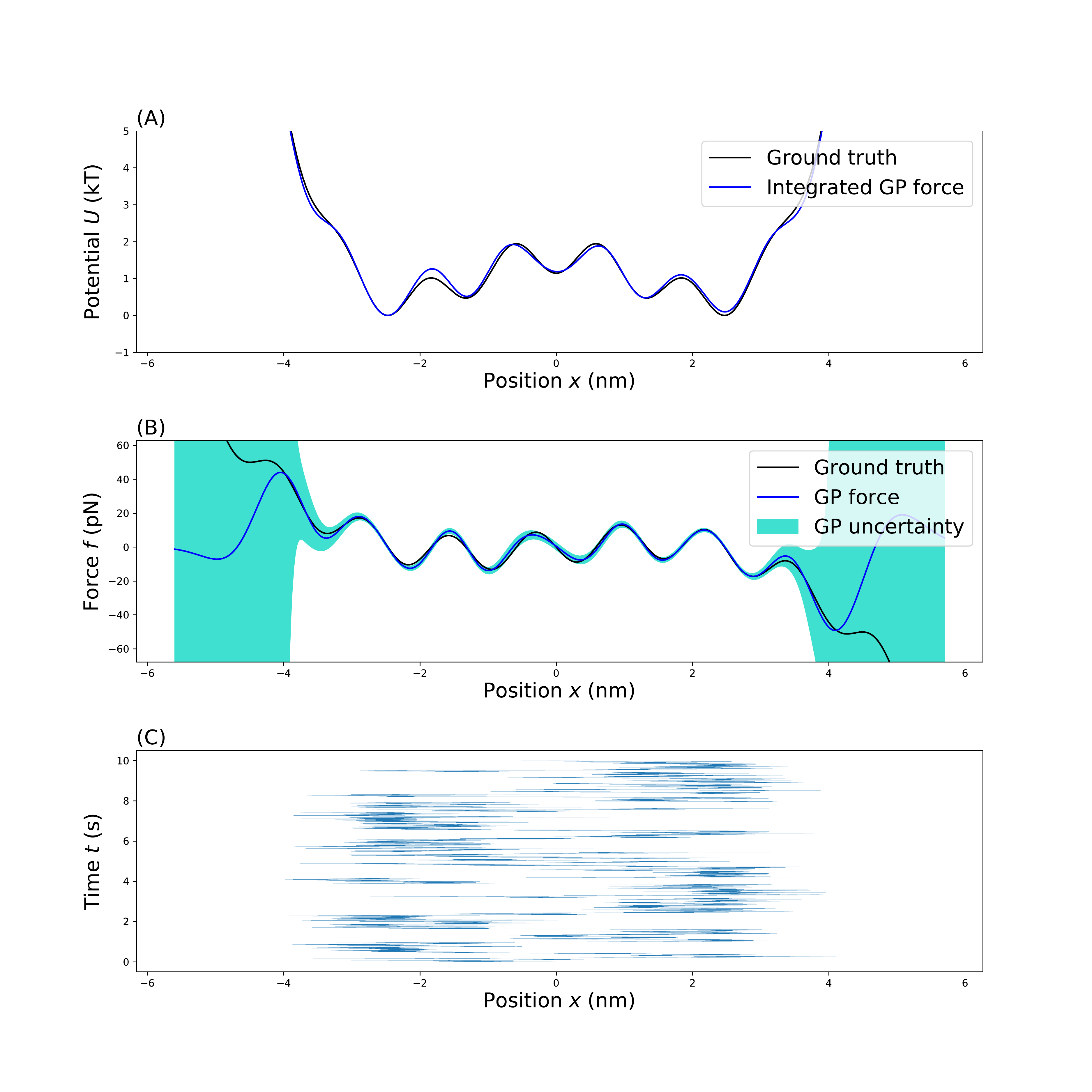}
\caption{\textbf{Testing our method on a more complex, multi-well, potentials}. Here we show the effective force (and effective potential) inferred from the time trace of a particle diffusing within a multi-well potential. (A) we show the ground truth and inferred effective potential. (B) the ground truth force and the MAP estimate for the inferred effective force with uncertainty. (C) the time trace on which the inference was performed.}
\label{fig:simplepic2}
\end{figure*}

\subsubsection{Application to a complicated potential}

We then tested the robustness of our method on a more complicated (multi-well) effective potential. We chose a spatially-varying, complicated potential to test how well our method can pick up on finer, often undersampled, potential details like potential barriers. To better pick up on fine details, we decreased the length scale, $\ell$, appearing as a hyperparameter to $\ell = (x_{max}-x_{min})/10$ so that correlations between distant points die out faster (See SI). The results are shown in Fig.~\ref{fig:simplepic2}. For a trace with $N=10^4$ time points, our method was able to uncover all potential wells and barrier shapes with minimal uncertainty.

\subsubsection{Varying the stiffness}

Next, to test the robustness of our method with respect to the stiffness, we applied our method to data simulated with different values of $\zeta/\tau$.

\begin{figure*}[tbp]
\centering	
\includegraphics[width=0.9\textwidth]{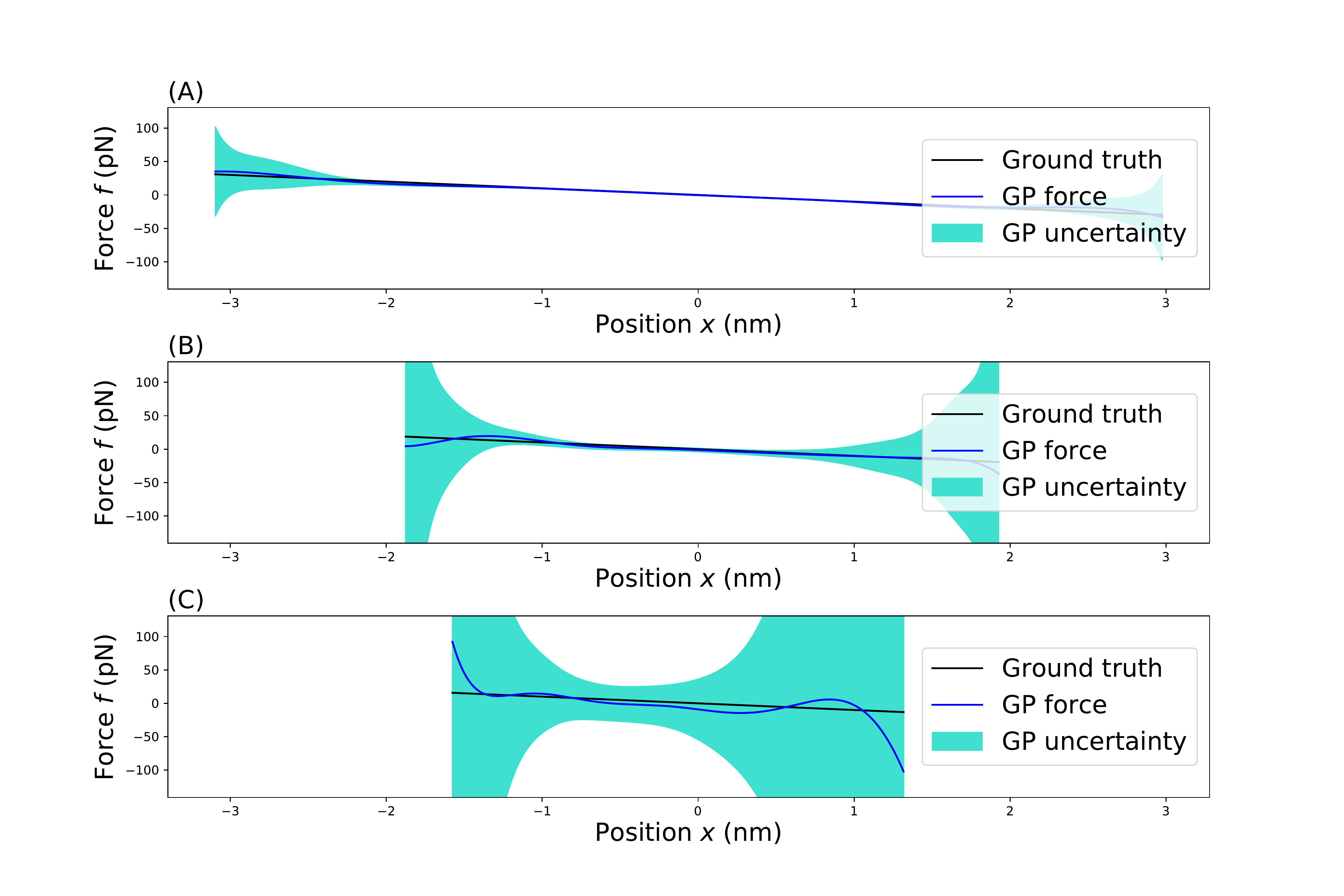}
\caption{\textbf{Testing our method for different stiffness}. Here we show the effective force inferred from 1000 point trajectories simulated for particles trapped in a one dimensional harmonic well using different choices of stiffness. (A) Stiffness is $h = 100~\mbox{pg/}\mu\mbox{s}$. (B) Stiffness is $h = 1000~\mbox{pg/}\mu\mbox{s}$. (C) Stiffness is $h = 10000~\mbox{pg/}\mu\mbox{s}$.}
\label{fig:gammas1}
\end{figure*}

Re-writing our equation with a new parameter, $h = \zeta/\tau$ called the stiffness,
\begin{align}
h(x_{n+1}-x_n) &= f\left( x_n \right) + r_{n}
\\
r_{n}&\sim\normal\left(0,2hkT\right).
\end{align}
we see that varying the stiffness can be interpreted as changing the friction coefficient or the capture rate of the time trace, 
as $\zeta$ and $\tau$ always appear together in our model. The results are shown in Fig.~\ref{fig:gammas1}.

For large values, the MAP estimate captures some of the trend, but the uncertainty remains large as thermal kicks at each time step dominate the dynamics.~\citet{Frishman2018} describe this in analogy to a signal to noise ratio. For small values of $h$ the prediction is, predictably, dramatically improved.

\subsubsection{Gibbs sampling for friction coefficient}

\begin{figure*}[tbp]
\centering	
\includegraphics[width=0.9\textwidth]{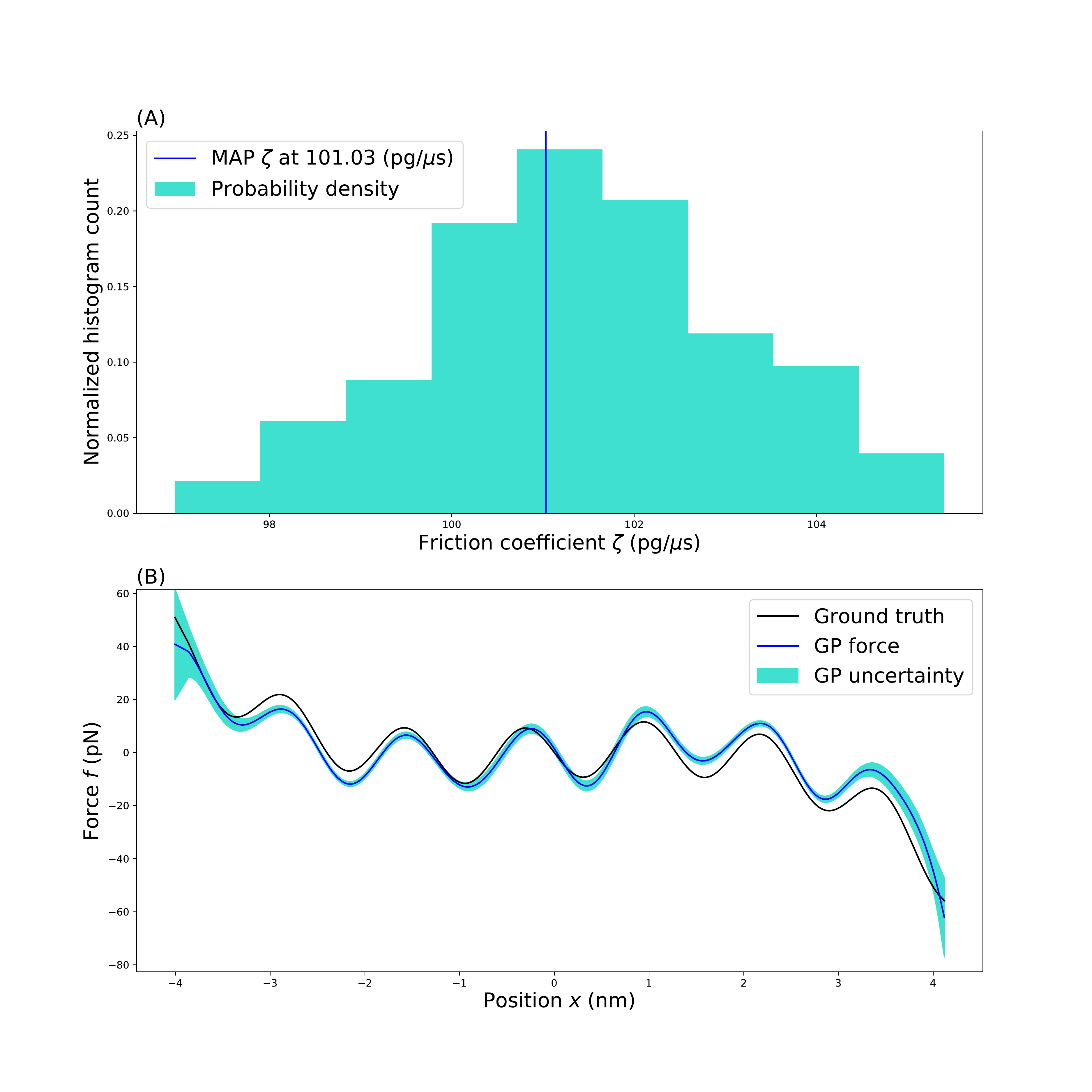}
\caption{\textbf{Results of simultaneous force and friction coefficient determination using Gibbs sampling}. This figure shows the inference of the force and friction coefficient using a Gibbs sampler. (A) A histogram of the sampled friction coefficients simulates the posterior on $\zeta$. The highest probability sampled $\zeta$ was $101.03~\mbox{pg/}\mu\mbox{s}$. (B) The MAP force matches nicely with the ground truth force.}
\label{fig:Gibbs1}
\end{figure*}

We used a Gibbs sampling scheme to simultaneously sample a complicated force and its corresponding friction coefficient from the time trace of position versus time. For our prior on $\zeta$ we used a shape parameter $\alpha = 1$ and a scale parameter $\beta = 1000~\mbox{pg/}\mu\mbox{s}$ so that we could allow $\zeta$ to be sampled over a large range. By histogramming the sampled $\zeta$'s, we simulated the posterior on the friction coefficient (Fig.~\ref{fig:Gibbs1} A). Here we see that the sampled values converged close to the true value. While not as good as the prediction with known friction coefficient (as also determining the friction coefficient demands more from the finite data available) we saw earlier, 
the prediction is still satisfactory. This demonstrates that the effective force can still be learned with an unknown friction coefficient.

\subsection{Comparison with other methods}

We compare our method described above to two previous methods to which we can compare most directly: that of  Masson {\it et al.}~\cite{Masson2009} and the residence time method~\cite{reif2009fundamentals}. Because the main proof of principle is for learning the force, we compare all methods for known friction coefficient. In order to compare our method to other methods, we use the following metric: how much data it takes to converge to a force that is reasonably close to the ground truth. This will be shown in Figs.~\ref{fig:comparison} and~\ref{fig:AccuracyVsN}.

\subsubsection{Masson {\it et al.}}

We first compare to Masson {\it et al.}'s method~\cite{Masson2009} which follows a Bayesian strategy similar in spirit to ours. This method uses the same likelihood as in Eq.~\eqref{likelihoodlang}, but spatially bins the data into $S$ equally sized bins and uses flat priors on the force in each bin instead of the GP prior of Eq.~\eqref{gpprior}. This approach then finds a force and friction coefficient which maximizes the posterior~\cite{Masson2009}. This method is optimized for simultaneously learning force and friction coefficient, but for simplified comparison, we hold $\zeta$ fixed. For known friction coefficient and one dimensional motion, this method reduces to averaging the measured force in each bin. This method predicts the effective force strength in bin $s$ to be
\begin{align}
    f(x) &= \frac{1}{C_s}\sum_{n: x_n\in s} \frac{\zeta}{\tau} (x_{n+1} - x_{n})
\end{align}
where $C_s$ is the number of times the position was measured in bin $s$. Masson {\it et al.} describe a method for choosing the number of bins, $S$, using the mean square displacement of the time trace, however for effective potentials which have drastic change over small ranges, this method does not give bin numbers that capture the rapid change. Therefore when we compare to this method we choose bin numbers by hand.

Looking at Fig.~\ref{fig:AccuracyVsN} A we see that our method converges to within a pre-specified error with fewer data points. The error was determined by numerically integrating the absolute value of the predicted effective force subtracted from the ground truth force over a range between $-1$~nm and $+1$~nm. We chose this range because it is the region with the most data and thus, the region in which the Masson {\it et al.} method best performs. For the Masson {\it et al.} method, smaller bin size led to a smaller error. However this effect emerged only after a large number of data points because for smaller bins, there are fewer data points per bin. Similarly, Masson {\it et al.}'s method also suffers in regions with few data points as this method averages the force from points within each bin. Outlier data points heavily influence effective forces inferred in these regions. On the right hand side of the harmonic force (Fig.~\ref{fig:comparison} A), we see that a few outlier measurements dominate the inference in this region. For this reason the method reflects a highly incorrect predicted effective force. 
On the other hand, within our method, the smoothness assumption imposed by the GP prior prevents the prediction from being influenced by these points too heavily. 

For the quartic potential (Fig.~\ref{fig:comparison} C), the Masson {\it et al.} method under-estimates the effective force in the steep parts of the effective potential. This is because the method averages all the points in the bin equally regardless of location. Because the force in each bin is not spatially constant, the sides of each bin will have different forces. There inevitably are more points at the low force side of each bin than at the high force side and so each bin is dominated by data points displaying a lower effective force. The GP prior avoids binning altogether and thus avoids this problem altogether.

\subsubsection{Residence time}

We also compared our method to the residence time (RT) analysis for finding the effective potential~\cite{Florin1998}. This method also bins space; but totally ignores times. For each bin, $s$, this method assumes that under Boltzmann statistics, the expected number of times a particle is found in bin $s$ after $N$ measurements is given by
\begin{align}
    N_s =N e^{-U_s/kT}.
\end{align}
Solving for $U_s$ this method predicts the effective potential in bin $s$ to be
\begin{align}
    U_s = -kT\ln(N_s/N).
\end{align}

\begin{figure*}[tbp]
\centering	
\includegraphics[width=.9\textwidth]{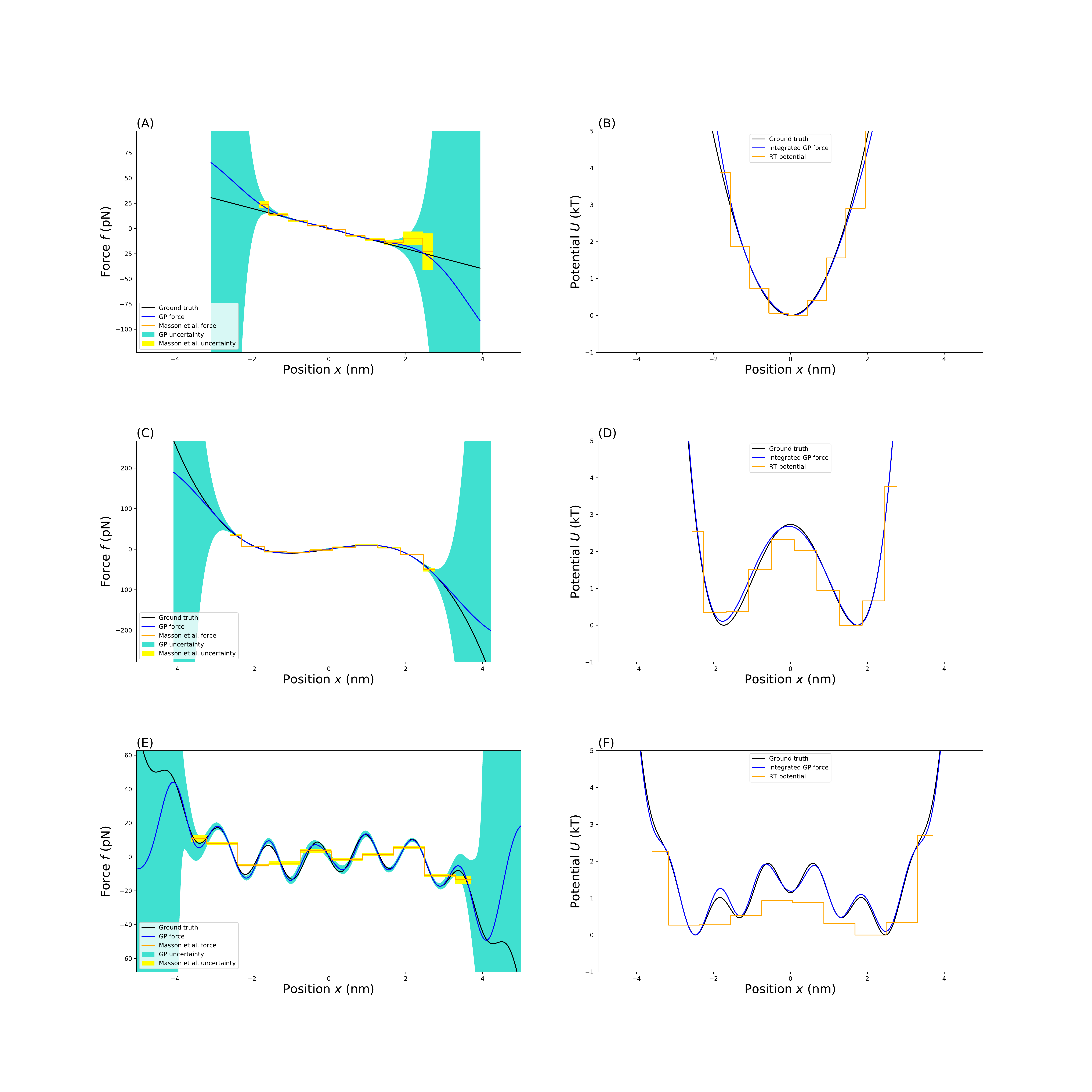}
\caption{\textbf{Comparison of our method with others}. In this figure, we compare our method to other methods of inference on three different effective force fields. The top compares inferences on the simplest, harmonic, potential  (A,B). The middle compares inference on a quartic potential (C,D). The bottom compares inference on a more complex potential (E,F). The left side of this figure compares our method of finding effective force with the Masson {\it et al.} method (A,C,E). The right side of this figure compares the integral of our inferred effective force field with the RT method for finding energies (B,D,F).}
\label{fig:comparison}
\end{figure*}

\begin{figure*}[tbp]
\begin{center}
\includegraphics[width=0.9\textwidth]{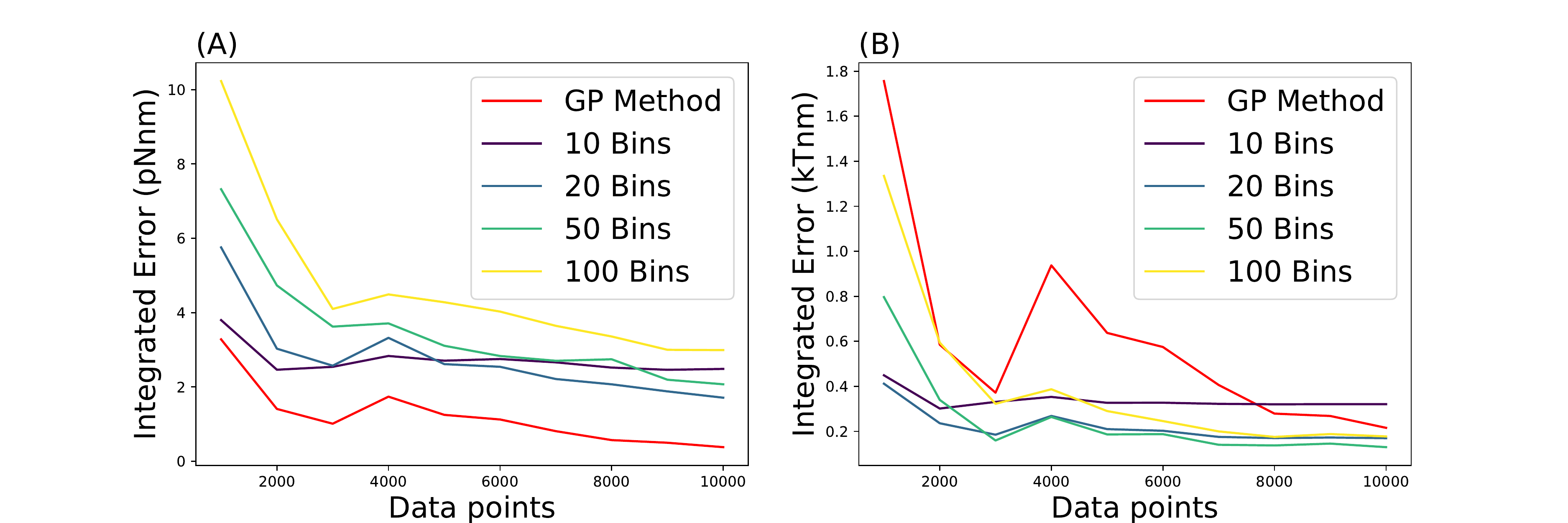}
\caption{\textbf{Error versus data points}. Here we compare the error of each method. Error was calculated by integrating the absolute value of the predicted effective force (A) or potential (B) subtracted from the ground truth force or potential over the range $[-1~\mbox{nm},1~\mbox{nm}]$. Different bin numbers of Masson {\it et al.} and residence time were compared. (A) we compare the amount of data points it takes for our method and the Masson {\it et al.} method to converge to an accurate effective force field for the harmonic well. (B) we compare the amount of data points it takes for our method and the RT method to converge to an accurate effective potential for the harmonic well.}
\label{fig:AccuracyVsN}
\end{center}
\end{figure*}

For a harmonic potential the RT method gives a good prediction (Fig.~\ref{fig:comparison} B). Even so, our integrated MAP effective force still gives an answer that is more accurate (Fig.~\ref{fig:AccuracyVsN} B). For the more complex, multi-well, potential (Fig.~\ref{fig:comparison} F) the prediction by RT is not satisfactory. Here the RT method underestimates the potential and does not capture all of the fine detail. For the quartic potential force, while our method predicts a highly accurate force, it does slightly overestimate the potential in the left well. That is because our method is tailored for accurate effective force predictions and potential predictions are a post processing step which introduces additional (albeit small and bounded) errors due to integration (see SI).

\section{Discussion}

Learning effective forces is an important task as knowledge of appropriate forces allows us to develop reduced models of dynamics of complex biological macromolecules~\cite{Wang2018,Poltavsky2017,Kumar1992,Foley2015, Izvekov2020, MARTINI_MD}. Here, we developed a framework in which we inferred effective force profiles from time traces of position versus time for which positions may very well be undersampled. In order to learn both friction coefficient and forces simultaneously from the same data set--and avoid binning or other pre-processing procedures that otherwise introduce modeling artifacts--we developed a Gibbs sampling scheme and exploited Gaussian process priors over forces. 

As such, our method simultaneously exploits every data point without binning or pre-processing and provides full Bayesian posterior over the prediction of the friction coefficient and force while placing minimal prior assumptions on the force profile. 

A limitation of our method is its poor scaleability. This is an inherent limitation of GP themselves, as these statistical tools naturally require the inversion of large matrices. The computations required to perform these inversions scale roughly as $(N+M)^3$ where $N,M$ are the number of datapoints and test points respectively and, as such, the memory required to do this for any data sets larger than roughly 50,000 data points is above the capability of an average desktop computer.
Moving forward, it is worth noting that approximate inference for GP methods are under active development~\cite{nemeth2018,Titsias2011}.

Beyond computational improvements which are necessary for large scale applications, there are many ways to generalize the framework we proposed.
In particular, at some computational cost, the form for the likelihood could be adapted to treat different models including measurement noise. Reminiscent of the hidden Markov model\cite{rabiner1986introduction,Rabiner}, this may be achieved by adapting of Gibbs sampling scheme to learn the positions themselves hidden by the noise from the data~\cite{Jazani2019}. Furthermore, just as we placed a prior on the friction coefficient, we may also place a prior on the initial condition or even potentially treat more general (non-constant) friction coefficients~\cite{Satija2017,Presse2013,Presse2014}.  

Yet another way to build off of our formulation is to generalize our analysis to multidimensional trajectories. Multidimensional data sets are common such as when combining single molecule measurements from force spectroscopy and F{\"o}rster resonance energy transfer~\cite{Sgouralis2018,Sgouralis2019,Jazani2019}. To treat these data sets within our GP framework is straightforward as we would simply need to choose a kernel allowing two- or higher-dimensional inputs\cite{Rasmussen:2005:GPM:1162254}.

In conclusion we have demonstrated a new method for finding the effective force field of an environment with minimal prior commitment to the form of the force field, exploiting every data point without binning or pre-processing, and which can predict the force on the full posterior with credible intervals.

\begin{acknowledgement}
The authors thank the NSF Award No.~1719537 ``NSF CAREER: Data-Driven Models for Biological Dynamics'' (Ioannis Sgouralis, force spectroscopy), the Molecular Imaging Corporation Endowment for their support, and NIH award No.~1RO1GM130745 ``A Bayesian nonparametric approach to superresolved tracking of multiple molecules in a living cell'' (Shepard Bryan, Bayesian nonparametrics).
\end{acknowledgement}

\subsection{Author Contributions} JSB analyzed data and developed analysis software; JSB, IS, SP conceived research; SP oversaw all aspects of the projects.

\subsection{Supplemental information}

\subsubsection{Probability distributions}

In this manuscript we use the Gaussian (Normal) and gamma distributions. The normal is defined as
\begin{align}
    \normal(x|\mu,\sigma^2) &= \frac{1}{\sqrt{2\pi\sigma^2}}e^{-\frac{1}{2\sigma^2}(x-\mu)^2}
\end{align}
with multivariate extension
\begin{align}
\normal_K(\mtx{x}|\mtx{\mu},\mtx{\Sigma}) &= \frac{1}{\sqrt{(2\pi)^K|\mtx{\Sigma}|}}\exp\left(-\frac{1}{2}(\mtx{x}-\mtx{\mu})^T\mtx{\Sigma}^{-1}(\mtx{x}-\mtx{\mu})\right).
\end{align}
In the multivariate normal, $\mtx{x}$ and $\mtx{\mu}$ are column vectors of size $K$ and $\mtx{\Sigma}$ is a square matrix of size $K\times K$.

The gamma distribution is defined as
\begin{align}
    \Gammapdf(x|\alpha,\beta) &= \frac{1}{\Gamma(\alpha)\beta^\alpha} x^{\alpha-1}e^{-x/\beta}
\end{align}
where $\Gamma(\alpha)$ is the Gamma function.

\subsubsection{Obtaining potential from force}

Given a conservative force, $f(\cdot)$, we can obtain its corresponding potential, $U(\cdot)$, using the relation $f = \nabla U$. In one dimension, this becomes $f=\frac{d}{dx} U$,
thus
\begin{align}
    U &= \int f dx + C.
\end{align}
In practice we numerically integrate $f(\cdot)$ over $\mtx{x}^*_{1:M}$ (assuming $\mtx{x}^*_{1:M}$ are not too sparcely distributed). This solves $U(\cdot)$ to an arbitrary constant. This constant can be any number; we choose that it be equal to the value of the minimum of $U(\cdot)$, specifically $C=-\min_{x^*_m}U(x^*_{m})$, so that
the deepest part of the potential well is defined to be zero.

\subsubsection{Bayes' theorem and the GP prior}
 
In our method we use Bayes' theorem
\begin{align}
    \prob\left(f(\cdot)|\mtx{x}_{1:N},\zeta \right) &\propto \prob\left(\mtx{x}_{1:N}|f(\cdot),\zeta \right) \prob\left(f(\cdot)\right)
\end{align}
where $\prob\left(f(\cdot)|\mtx{x}_{1:N},\zeta\right)$ is the posterior, $\prob\left(\mtx{x}_{1:N}|f(\cdot),\zeta\right)$ is the likelihood, and $\prob\left(f(\cdot)\right)$ is the prior. The best estimate for $f(\cdot)$ is the one maximizing the posterior. This is called the maximum a posteriori (MAP) estimate.

Rather than look at the continuous force $f(\cdot)$, we look at the effective force at $M$ arbitrarily chosen points, $\mtx{x}^*_{1:M}$. These points are called the test points. We stress that limiting our inference to these points does not limit our ability to infer the effective force on the full real line because the test points can be chosen anywhere we wish to know the effective force.

In order to proceed, we discretize our effective force. Let $\mtx{f}_{1:N}$ be the effective force at the $N-1$ data points, $f_n = f(x_n)$. We also introduce $M$ new points, $\mtx{x}^*_{1:M}$, where we wish to predict the effective force. Let $\mtx{f}^*_{1:M}$ be the effective force at the test points, $f^*_m = f(x^*_m)$. The discretized Gaussian process prior becomes
\begin{align}
\prob\left(\mtx{f}_{1:N-1}, \mtx{f}^*{_{1:M}}\right) &=
        \normal_{N+M-1}\left(\left[\begin{matrix} \mtx{f}_{1:N-1}\\ \mtx{f}^*{_{1:M}}\end{matrix} \right] ; \mtx{0}, \left[ \begin{matrix} \mtx{K} & \mtx{K_*} \\ \mtx{K_*}^T & \mtx{K_{**}} \end{matrix} \right]\right)\\
        K(x,x') &= \sigma^2 \exp\left(-\frac{1}{2} \left(\frac{x_i-x^*_j}{\ell}\right)^2 \right)\label{kernalfunc}
\end{align}
where $\mtx{K}$ is the covariance matrix between the data, $\mtx{K_{**}}$ is the covariance matrix between the test points, and  $\mtx{K_*}$ is the covariance matrix between the data points and the test points. The elements of these covariance matrices are given by the kernel function (Eq.~\eqref{kernalfunc}), for example $K_{*ij} = \sigma^2 \exp\left(-\frac{1}{2}\left(\frac{x_i-x^*_j}{\ell}\right)^2\right)$. This choice for prior imposes a smoothness assumption on the effective force by correlating nearby points. The effect of this assumption can be tuned by two hyperparameters, $\sigma$ and $\ell$, both of which attain positive scalar values.

\subsubsection{Hyperparameters}

\begin{figure*}[tbp]
\centering	
\includegraphics[width=0.9\textwidth]{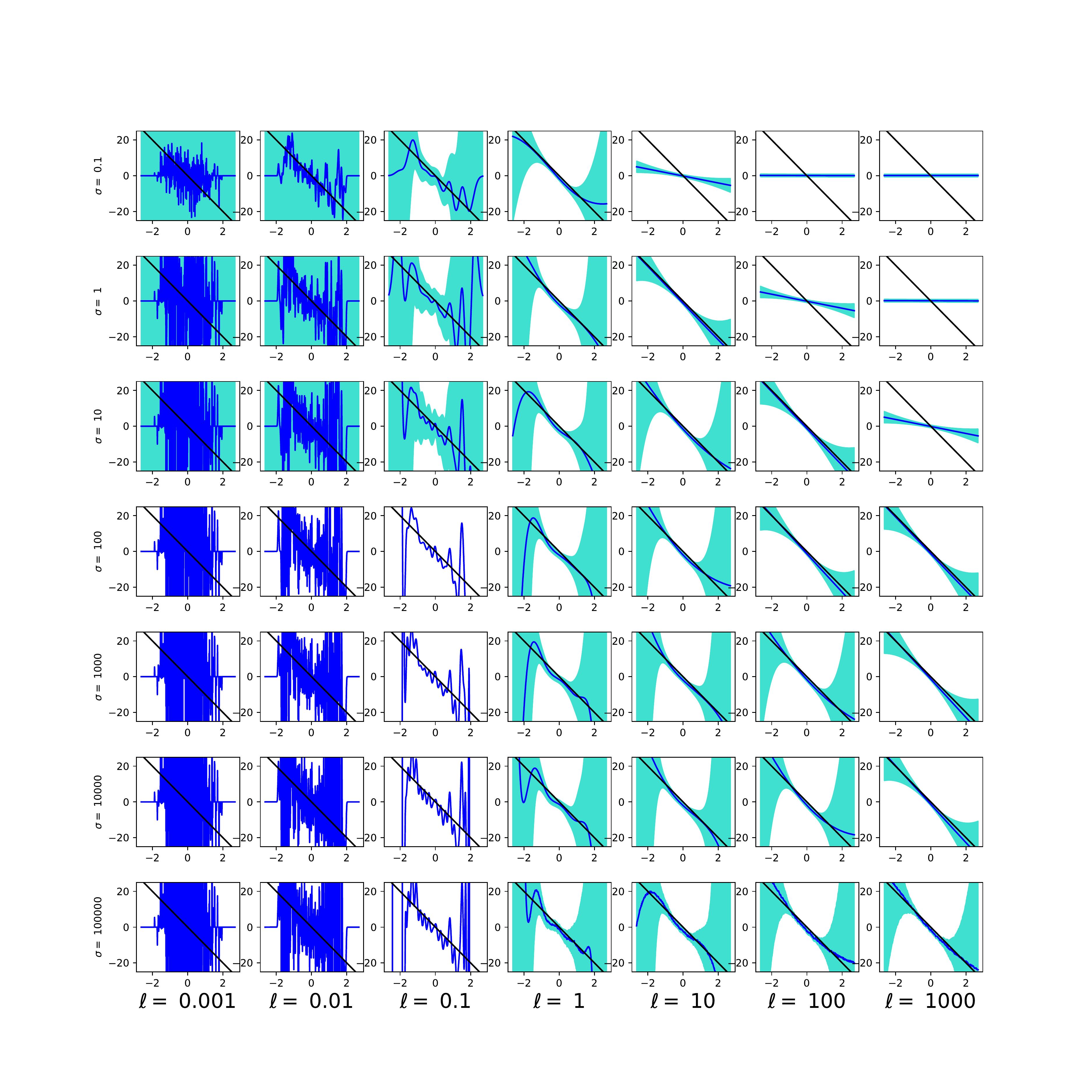}
\caption{\textbf{Varying the hyperparameters}. This plot shows inference of the effective force from a 1000 point time trace for a simulated particle trapped in a one dimensional harmonic well using different choices of hyperparameters. The prefactor, $\sigma$ is varied on the vertical axis and the length scale, $\ell$, is varied on the horizontal. For all plots in this figure black is the ground truth force, red is the MAP estimate, and dotted blue is the uncertainty.}
\label{fig:hyperparameters1}
\end{figure*}

The hyperparameters allow us to tune the prediction. Algorithms exist to optimize the hyperparameters~\cite{Rasmussen:2005:GPM:1162254}, but for our purpose it suffices to input them by hand. Fig.~\ref{fig:hyperparameters1} shows how different choices for hyperparameters lead to different predictions. For simplicity, the values for $\sigma$ shown in the figure represent that value times the range in the data. For example, $\sigma = 10$ in the figure corresponds to $\sigma = 10\alpha\tau( v_{max}- v_{min})$ (with $\alpha = 1~\mbox{pN}\mbox{/nm}$) in the inference. Likewise, $\ell=.1$ in the figure corresponds to $\ell = (x_{max}-x_{min})/10$ in the inference.

The first hyperparameter, $\sigma$, is called the prefactor. It determines the uncertainty in the regions with no data. A greater $\sigma$ value also pulls the prediction in closer to the measurements. The second hyperparameter, $\ell$, is called the length scale. It defines the range in which correlations die out. The smaller the length scale, the more rapid the variation we get for our predictions, however, a length scale too large means that the prediction will correlate strongly with points where there is no data.

\begin{figure*}[tbp]
\centering	
\includegraphics[width=0.9\textwidth]{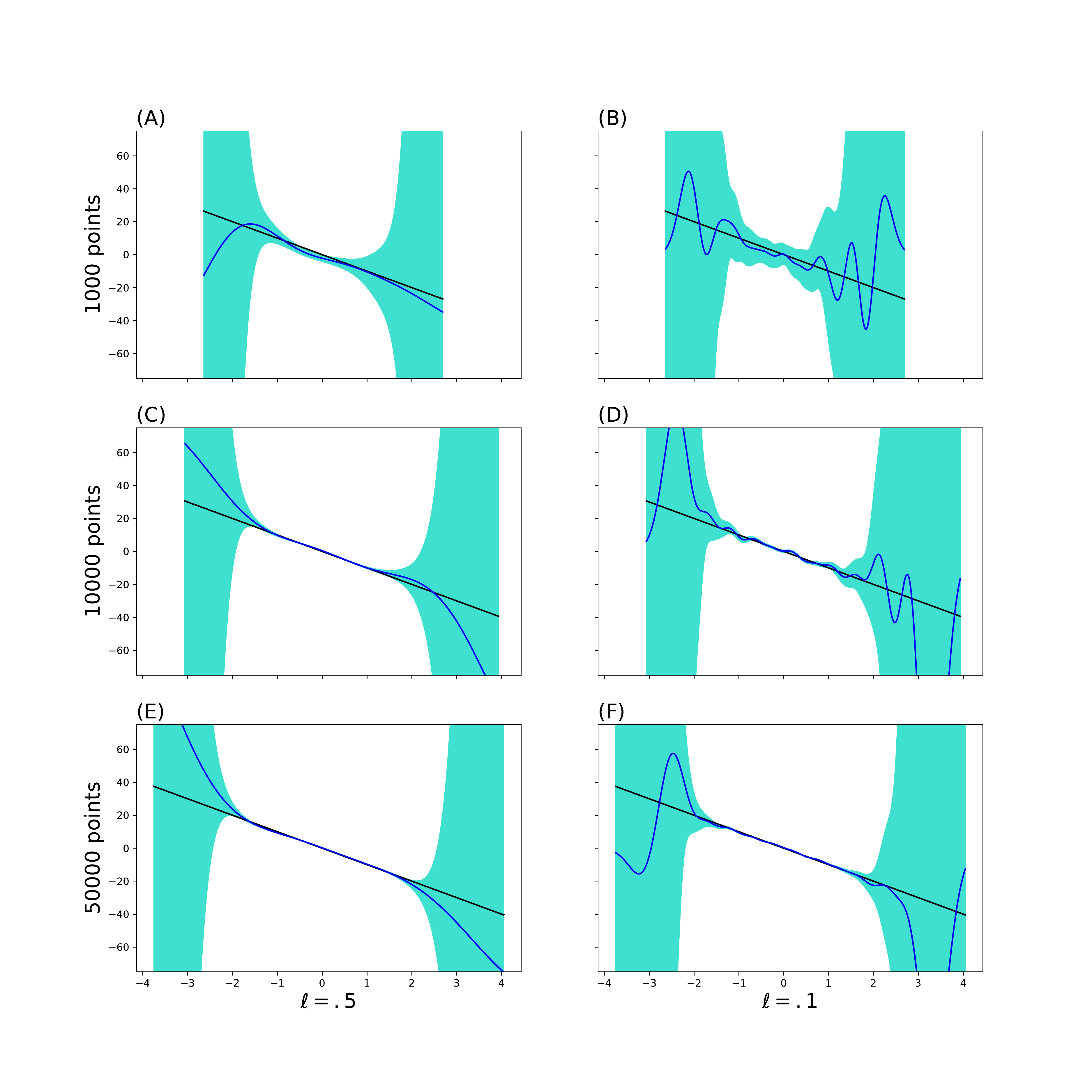}
\caption{\textbf{Length scale determines detail of prediction 1}. This plot shows the learned force from a low detail force field for different length scales and data points. Here the rows show different number of data points used and the columns show the different length scale used. The predicted force using a larger length scale is more accurate than the one from a small length scale.}
\label{fig:HypeN1}
\end{figure*}

\begin{figure*}[tbp]
\centering	
\includegraphics[width=0.9\textwidth]{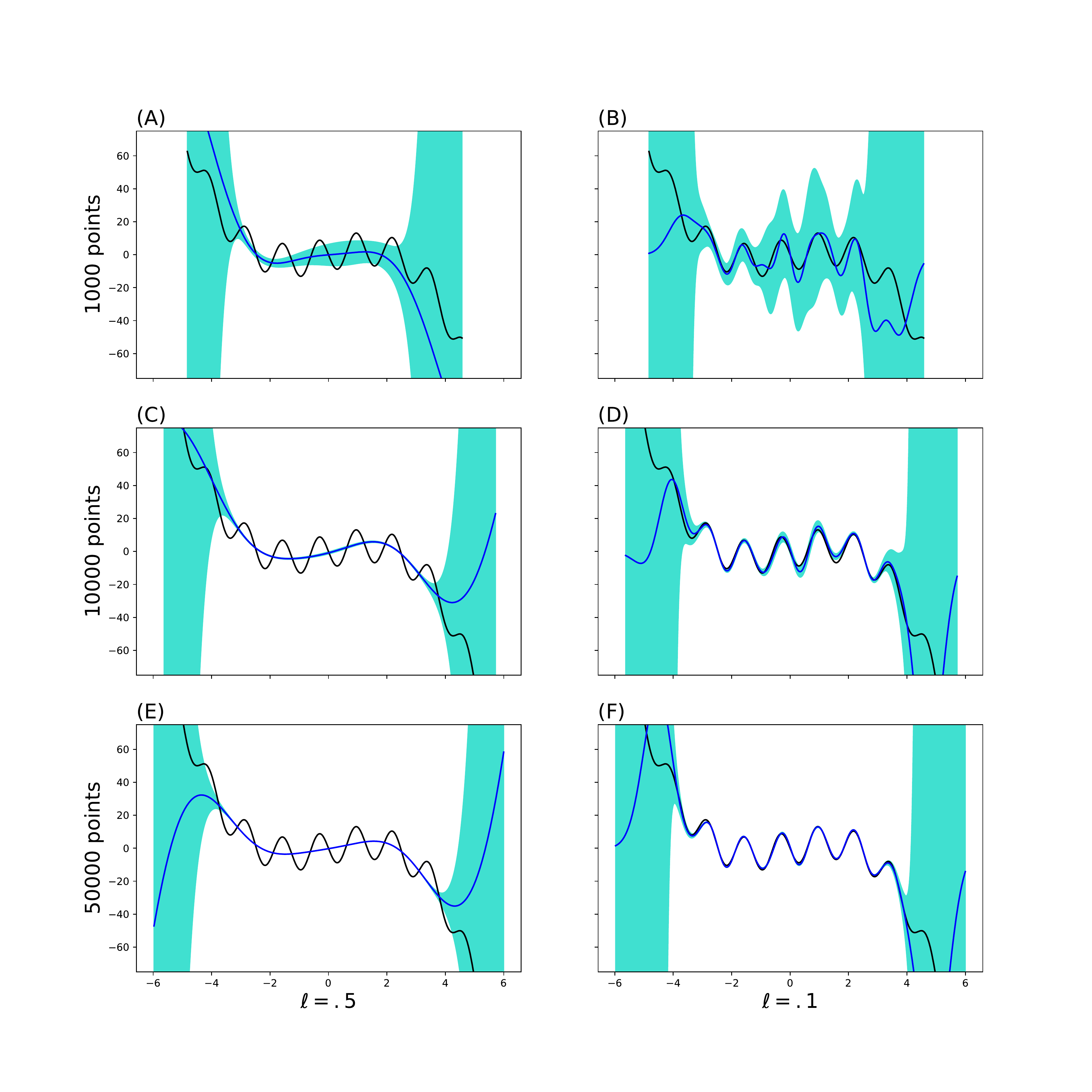}
\caption{\textbf{Length scale determines detail of prediction 2}. This plot shows the learned force from a high detail force field for different length scales and data points. Here the rows show different number of data points used and the columns show the different length scale used. The predicted force using a larger length scale does not pick up the fine details even at a high number of data points.}
\label{fig:HypeN2}
\end{figure*}

The length scale determines the level of detail desired for a prediction. For a effective force field with minimal fine detail, predictions with a larger length scale more accurately predicts the force (Fig.~\ref{fig:HypeN1}). The prediction using a small length scale picks up noise which decreases with more data points. For an effective force field with high level of detail, a prediction made using a large length scale will only pick up the large detail (Fig.~\ref{fig:HypeN2}). With more data points, this may eventually pick up the small detail, but because the GP has scaling problems, we could not try past 50,000 data points.

\bibliography{main}

\end{document}